# Fluctuation-Enhanced Sensing of Molecules with Biased Carbon Nanotubes, or Nanowires, and Adsorption-Desorption Fluctuations

Laszlo B. Kish[(1)], Robert Vajtai[(2)], Pulickel Ajayan[(2)], Chiman Kwan[(3)], Gabor Schmera[(4)]

[(1)] *Department of Electrical and Computer Engineering, Texas A&M University, College, Station, TX 77843-3128, USA*

[(2)] *Department of Material Science and Engineering, Rensselaer Polytechnic Institute, Troy, NY 12180, USA.*

[(3)] *Signal Processing, Inc., 13619 Valley Oak Circle, Rockville, MD 20850, USA*

[(4)] *Space and Naval Warfare System Center, Signal Exploitation & Information Management, San Diego, CA 92152-5001, USA*

**Abstract.** A new nanoscale device for fluctuation-enhanced sensing is proposed.

Fluctuation-Enhanced sensing (FES) [1] has recently been proposed for the detection of chemical and biological agents with sensitivity and selectivity superior to sensing methods based on the steady-state sensor signal. The stochastic component of the sensor signal is separated, strongly amplified and statistically analyzed to generate a pattern which serves as the finger print of the agent.

To utilize all the advantages of FES, the characteristic size of the sensor should be in the nanoscale range (non-Gaussian fluctuations) and a large number of micro-integrated integrated detectors (electronic dog nose) should be used for enhanced sensitivity and very low agent concentrations.

In this short note, we propose a simple FES element to generate a FES-based fingerprint of large molecules, see Figure 1. A carbon nanotube or nanowire pair, or ladder is deposited over a gateless MOSFET channel (covered with thin oxide). The pair is driven by a variable DC voltage generator and the ladder is driven by alternating polarity. Large agent molecules with inherent or induced polarization will be captured by the DC field which will influence the activation energy of desorption (evaporation) of the molecule from this site. The value of the activation energy E is characteristic of the molecule and even more information can be extracted from the function E(U), the dependence of the activation energy on the DC voltage. The presence of the molecule at the nanotube structure can be detected by the drain current of the MOSFET and the absorption-desorption process will generate a Lorentzian type excess noise, which can be observed provided the inherent noise of the MOSFET is low enough, and its characteristic crossover frequency provides information about the value of the activation energy.

The cleaning of chemical sensors is an especially difficult issue and the usual methods based overheating cause early aging and is not always sufficient. The present structure can however be cleaned very easily without heating by a short voltage pulse of opposite polarity. The pulse should be shorter than the polarization relaxation time of the molecule.

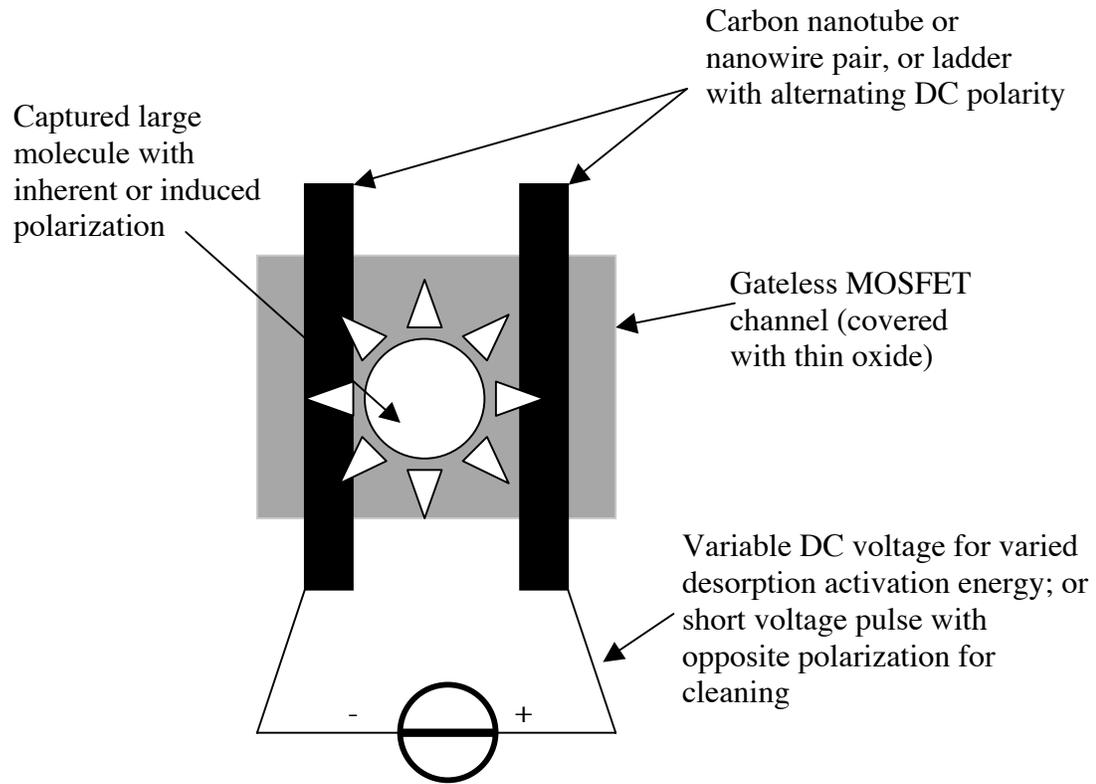

**Figure 1.** Outline of the FES element.